\documentclass[11pt]{article}
\usepackage{amssymb,latexsym,amsmath}
\usepackage[dvips]{graphicx}
\usepackage{cite}
\newtheorem{theo}{Theorem}

\newtheorem{prop}[theo]{Proposition}
\def\nn{\nonumber}
\def\deg{\mathop{\rm deg}\nolimits}

\def\qdots{\mathinner{\mkern1mu\raise1pt\vbox{\kern7pt\hbox{.}}\mkern2mu
 \raise4pt\hbox{.}\mkern2mu\raise7pt\hbox{.}\mkern1mu}}

\def\gl{\mathfrak{gl}}
\def\ssl{\mathfrak{sl}}
\def\u{\mathfrak{u}}

\def\osp{\mathfrak{osp}}
\def\lb{[\![}
\def\rb{]\!]}
\newcommand{\q}{\hat q}
\newcommand{\p}{\hat p}

\begin{document}
\begin{center}
{\Large \bf
Harmonic oscillator chains as Wigner Quantum Systems:\\[2mm]
periodic and fixed wall boundary conditions in $\gl(1|n)$ solutions. 
}\\[5mm]
{\bf S.~Lievens\footnote{E-mail: Stijn.Lievens@UGent.be}, }
{\bf N.I.~Stoilova}\footnote{E-mail: Neli.Stoilova@UGent.be; Permanent address:
Institute for Nuclear Research and Nuclear Energy, Boul.\ Tsarigradsko Chaussee 72,
1784 Sofia, Bulgaria} {\bf and J.\ Van der Jeugt}\footnote{E-mail:
Joris.VanderJeugt@UGent.be}\\[1mm]
Department of Applied Mathematics and Computer Science,
Ghent University,\\
Krijgslaan 281-S9, B-9000 Gent, Belgium.
\end{center}


\begin{abstract}
We describe a quantum system consisting of a one-dimensional linear chain
of $n$ identical harmonic oscillators coupled by a nearest neighbor interaction.
Two boundary conditions are taken into account: periodic boundary conditions (where
the $n$th oscillator is coupled back to the first oscillator) and fixed wall
boundary conditions (where the first oscillator and the $n$th oscillator are
coupled to a fixed wall). 
The two systems are characterized by their Hamiltonian.
For their quantization, we treat these systems as Wigner Quantum Systems (WQS), allowing
more solutions than just the canonical quantization solution.
In this WQS approach, one is led to certain algebraic relations for operators
(which are linear combinations of position and momentum operators) that should
satisfy triple relations involving commutators and anti-commutators.
These triple relations have a solution in terms of the Lie superalgebra $\gl(1|n)$.
We study a particular class of $\gl(1|n)$ representations $V(p)$, the so-called ladder representations.
For these representations, we determine the spectrum of the Hamiltonian and of the position
operators (for both types of boundary conditions).
Furthermore, we compute the eigenvectors of the position operators in terms of stationary
states. This leads to explicit expressions for position probabilities of the $n$
oscillators in the chain. 
An analysis of the plots of such position probability distributions gives rise to some
interesting observations. In particular, the physical behavior of the system as a WQS
is very much in agreement with what one would expect from the classical case, except that
all physical quantities (energy, position and momentum of each oscillator) have a 
finite spectrum.
\end{abstract}

\vfill\eject

%

\setcounter{equation}{0}
\section{Introduction}

Coupled systems describing interaction of oscillating or scattering subsystems and the
corresponding operators have been widely used in classical and
quantum mechanics. 
For example, coupled harmonic oscillators have been studied in quantum information theory, 
quantum optics (photonic crystals), and for describing phonons in a 
crystal~\cite{Cohen,Brun,Audenaert,Eisert,Halliwell,Plenio}.

In a previous paper~\cite{LSV06}, we initiated the analysis of such systems as a
Wigner Quantum System (WQS)~\cite{Wigner,Palev86,Palev1}.
The system studied in~\cite{LSV06} consists of a string or chain of $n$ identical harmonic
oscillators, each having the same mass $m$ and frequency $\omega$. 
The position and momentum operator for the $r$th oscillator ($r=1,2,\ldots,n$)
are given by $\hat q_r$ and $\hat p_r$; more precisely $\hat q_r$ measures the
displacement of the $r$th mass point with respect to its equilibrium position.
The oscillators are coupled by some nearest neighbor coupling,
represented by terms of the form $(\hat q_r - \hat q_{r+1})^2$ in the Hamiltonian.
In~\cite{LSV06}, we considered such a system with periodic boundary conditions,
where the last oscillator is again coupled to the first one (i.e.\ $\hat q_{n+1}= \hat q_1$).
In other words, the oscillators are on a circle rather than on a line.
The Hamiltonian of this system will be denoted by $\hat H_P$.
In the present paper we will reconsider this system, and at the same time also
concentrate on a second type of boundary conditions: fixed wall boundary conditions.
For this second case, the first and last oscillator are coupled to a fixed wall
(i.e.\ $\hat q_0 = \hat q_{n+1} =0$); the Hamiltonian of this second system
will be denoted by $\hat H_{FW}$.

The original contribution of~\cite{LSV06} consisted of the treatment of the
system described by $\hat H_P$ as a Wigner Quantum System~\cite{Wigner,Palev86,Palev1}.
This quantization procedure is based upon the compatibility of Hamilton's equations
and the Heisenberg equations. 
In this procedure, one does not require the canonical commutation relations (CCRs),
but instead uses more general relations describing the before mentioned compatibility.
As a consequence, one obtains different classes of solutions for the system, of which
the canonical solution is only one.
For a more detailed description of WQSs, see~\cite{LSV06} and references therein.
The main feature of this treatment is that the CCRs are replaced by these
compatibility conditions (CCs). For the Hamiltonian $\hat H_P$, these CCs
are triple relations in terms of operators $a_r^\pm$ ($r=1,2,\ldots,n$),
involving both commutators and anti-commutators. Herein, the operators $a_r^\pm$ 
are certain linear combinations of the position and momentum operators.
In~\cite{LSV06} we showed that these triple relations have a solution
in terms of certain generators of the Lie superalgebra $\gl(1|n)$~\cite{Kac1}. 
Furthermore, we considered a class of Fock representations~\cite{Palev2} of $\gl(1|n)$ and
analysed the energy spectrum and the position operator spectrum in these
Fock representations $W(p)$. These Fock representations are rather restricted, however,
and do not illustrate the features of general unitary irreducible representations~\cite{King}
of $\gl(1|n)$ (more precisely, of its compact form $\u(1|n)$).

In the present paper, we consider a more general class of $\gl(1|n)$ representations,
the so-called ladder representations $V(p)$~\cite{King}. 
These representations are also easy to describe, but more importantly they show 
interesting properties of the physical quantities (energy spectrum,
position operator spectrum) of the system, far more general than those of the Fock representations $W(p)$.

A second original contribution of this paper is that we show how also the second system (fixed wall
boundary conditions) described
by $\hat H_{FW}$ can be treated as a Wigner Quantum System. 
In fact, we show that at the algebraic level the two Hamiltonians give rise to the
same triple relations, but with different constants. 
As a consequence, the two systems can be treated similarly, both having solutions
in terms of the Lie superalgebra $\gl(1|n)$.

The contents of the paper is as follows. In Section~2 we shortly review the treatment of $\hat H_P$
as a WQS~\cite{LSV06}, and then analyse the Hamiltonian $\hat H_{FW}$ of the fixed wall boundary
case in a similar way. As the algebraic triple relations are, up to different constants, the same,
we describe the $\gl(1|n)$ solution in a subsection. Special attention is paid to determining the
critical value for the coupling constant $c$, for which a $\gl(1|n)$ solution exists in the
fixed wall boundary case. We also describe briefly the new class of representations $V(p)$, 
the ladder representations, and the explicit action of $\gl(1|n)$ generators on simple basis
vectors $w(\theta;s)$ of $V(p)$. 
In Section~3 we determine the spectrum of the Hamiltonians $\hat H_P$ and $\hat H_{FW}$ in
the representations $V(p)$. Due to the fact that these Hamiltonians have a diagonal action
in the current model, their spectrum is easy to determine. We compare the two cases, and
discuss some aspects of degeneracy of the energy levels.

The mathematically more difficult problem is that of determining the eigenvalues and eigenvectors
of the position operators $\hat q_r$, for these representations $V(p)$, since $\hat q_r$ is
a general odd operator of $\gl(1|n)$ with a nondiagonal action on the standard basis of $V(p)$.
Fortunately, we have managed to give a general treatment (and procedure) of how to construct such
eigenvalue spectrum and eigenvectors for an arbitrary odd element of $\gl(1|n)$ in an arbitrary
unitary representation in~\cite{LSV07}.
Using the techniques of~\cite{LSV07}, we manage to determine the eigenvalues of $\hat q_r$
for both systems in Section~4, and to construct the eigenvectors of $\hat q_r$ in Section~5.
In fact, the eigenvalues of $\hat q_r$ for the system described by $\hat H_P$ were already given
in~\cite{LSV07} as an example of the new technique, so only those for the second system described by $\hat H_{FW}$
are new. We also give some plots of the spectrum of the position operators, as a function
of the coupling constant~$c$, and make some observations related to these plots.

Section~5 is devoted to computing position probability distributions for the two systems
under consideration.
Since the techniques of~\cite{LSV07} allow us to give an explicit expansion of the
normalized eigenvectors of $\hat q_r$, for any of its eigenvalues $\pm x_K$, in terms
of the stationary states $w(\theta;s)$, we can invert these relations and
express the stationary states $w(\theta;s)$ in terms of the eigenvectors of $\hat q_r$
(for any particular $r$). 
The square moduli of the coefficients in these expressions have the usual quantum theory
interpretation as the probability of finding the $r$th oscillator in position $\pm x_K$
when the system is in its stationary state $w(\theta;s)$. 
We compute these probabilities analytically (for small $p$-values), and numerically
for some examples. We also give a number of plots of such probability distribution
functions. These yield the ``spacial properties'' of the oscillators.
The results are quite interesting, and show some analogy of what one would expect 
of the system from a classical treatment.
For the ground state (stationary state of lowest energy), the position with highest probability
(for each oscillator)
is the equilibrium position (corresponding to $x_p=0$), with strongly decreasing probability as
the (discrete) eigenvalue is further away from 0. For the most excited state (stationary state of highest
energy), the position with highest probability is away from 0 (symmetrically to the left and
the right). The higher the coupling constant, the further away this highest probability position.
The probability of the equilibrium position is zero here. All oscillators have the same position
probability distribution in the periodic boundary case, as they are all equivalent (completely symmetric
on a circle).
In the case of fixed wall boundary conditions, the behavior also follows the classical properties,
with the first and last oscillators (those fixed to the wall) ``oscillating less'' compared to those
away from the wall.
We end the paper by some concluding remarks in Section~6.

\setcounter{equation}{0}
\section{The Hamiltonian of the system: solutions as a WQS}

\subsection{System with periodic boundary conditions}

In this subsection, we briefly recall the algebraic treatment of the Hamiltonian
for a system consisting of coupled harmonic oscillators, with periodic boundary conditions,
treated as a Wigner Quantum System~\cite{LSV06}. 
In several models~\cite{Brun,Audenaert,Eisert,Halliwell,Plenio} such a quantum system consisting of 
a linear chain of $n$ identical harmonic oscillators
coupled by springs is used. The Hamiltonian of such a system is given by:
\begin{equation}
\hat{H}_P=\sum_{r=1}^{n} \Big( \frac{\hat{p}_r^2}{2m}
+ \frac{m\omega^2}{2} \hat{q}_r^2 + \frac{cm}{2}(\hat{q}_r-\hat{q}_{r+1})^2  \Big),
\label{H_P}
\end{equation}
where each oscillator has mass $m$ and frequency $\omega$, $\q_r$ and $\p_r$ stand
for the position and momentum operator for the $r$th oscillator (or rather,
$\q_r$ measures the displacement of the $r$th mass point with respect to its
equilibrium position), and $c>0$ is the coupling strength. In the case of
periodic boundary conditions (indicated by the index in $\hat{H}_P$), one assumes in~\eqref{H_P}
\begin{equation}
\q_{n+1}\equiv \q_1.
\label{qn+1}
\end{equation}
With these periodic boundary conditions one can think of the oscillators as being located
on a circle, as the last oscillator is again coupled to the first one.

In~\cite{LSV06}, it was shown that one can relax the canonical commutation relations 
for the operators $\q_r$ and $\p_r$, leading to a larger class of solutions for the system
described by~\eqref{H_P}. This is known as
a Wigner Quantum System approach~\cite{Palev86,Palev1}. In this approach, one imposes 
the compatibility of Hamilton's equations
\begin{equation}
    {\dot{\hat{q}}}_r=\frac{\partial \hat H_P}{\partial \hat{p}_r}, \quad 
		{\dot{\hat{p}}}_r=-\frac{\partial \hat H_P}{\partial \hat{q}_r} \qquad (r=1,2,\ldots,n)
     \label{Ham}
\end{equation}
(formal derivatives) and the Heisenberg equations
\begin{equation}
     {\dot{\hat{p}}}_r = \frac{i}{\hbar}[\hat{H}_P,\hat{p}_r], \quad
     {\dot{\hat{q}}}_r = \frac{i}{\hbar}[\hat{H}_P,\hat{q}_r]  \qquad 
     (r=1,2,\ldots,n)
     \label{Heis}
\end{equation}
when viewed as operator equations.  These compatibility conditions (CCs) read
\begin{align}
  [\hat{H}_P,\hat{q}_r] &= -\frac{i\hbar}{ m} \hat{p}_r , \label{CCpq1}\\{} 
  [\hat{H}_P,\hat{p}_r] &=-i\hbar cm\,\q_{r-1}+i\hbar m(\omega^2+2c)\,\q_r-i\hbar cm \,\q_{r+1}, \label{CCpq2}
\end{align}
with $r\in\{1,2,\ldots,n\}$, $\hat q_{n+1} = \hat q_1$ and $\hat q_{0} = \hat q_{n}$. 
Under the canonical commutation relations,
the CCs~\eqref{CCpq1} and~\eqref{CCpq2} are automatically satisfied. But the system~\eqref{CCpq1}-\eqref{CCpq2},
with $\hat H_P$ given by~\eqref{H_P}, has other interesting solutions~\cite{LSV06}. 

To study these other solutions, one introduces discrete Fourier transforms of the
(self-adjoint) operators $\q_r$ and $\p_r$ by
\begin{align}
\q_r & = \sum_{j=1}^n \sqrt{\frac{\hbar}{2m n\omega_j}} \left( e^{-2\pi i jr/n} a_j^+ +
e^{2\pi i jr/n} a_j^- \right), \label{qr1}\\
\p_r & = \sum_{j=1}^n i\; \sqrt{\frac{m\omega_j\hbar}{2n}} \left( e^{-2\pi i jr/n} a_j^+ -
e^{2\pi i jr/n} a_j^- \right), \label{pr1}
\end{align}
where $\omega_j$ are positive numbers with 
\begin{equation}
\omega_j^2=\omega^2 +2c-2c\cos(\frac{2\pi j}{n})=\omega^2+4c\sin^2(\frac{\pi j}{n}),
\label{omega}
\end{equation}
and $a_j^\pm$ are operators satisfying $(a_j^\pm)^\dagger = a_j^\mp$. In terms of these new
operators, the Hamiltonian reads~\cite{LSV06}
\begin{equation}
\hat{H}_P=\sum_{j=1}^{n} \frac{\hbar \omega_j}{2} (a_j^- a_j^+ + a_j^+ a_j^-). 
\label{H1aa}
\end{equation}
Note that we no longer require the canonical commutation relations for the operators $\q_r$ and $\p_r$,
so also the operators $a_j^\pm$ no longer satisfy the usual boson relations 
$[a_j^\pm,a_k^\pm]=0$ and $[a_j^-,a_k^+]=\delta_{jk}$.
In the WQS approach, the relations that should be satisfied follow from~\eqref{CCpq1}-\eqref{CCpq2},
and read explicitly~\cite{LSV06}:
\begin{equation}
\Bigl[ \sum_{j=1}^{n}  \omega_j (a_j^- a_j^+ + a_j^+ a_j^-) , a_k^\pm \Bigr]=
\pm 2 \omega_k a_k^\pm, \quad \ (k=1,2,\ldots,n). 
\label{algrelations1}
\end{equation}

\subsection{System with fixed wall boundary conditions}
In this subsection we consider a similar system, again consisting of
coupled harmonic oscillators, but this time with fixed wall boundary conditions.
Now the Hamiltonian reads:
\begin{equation}
\label{H_FW}
\hat{H}_{FW}=\sum_{r=1}^{n} \Big( \frac{\hat{p}_r^2}{2m}
+ \frac{m\omega^2}{2} \hat{q}_r^2 + \frac{cm}{2}(\hat{q}_r-\hat{q}_{r+1})^2  \Big),
\end{equation}
with the same data as in~\eqref{H_P}, but 
\begin{equation}\label{BC2}
\hat q_0 = \hat q_{n+1} \equiv 0\quad\text{(and\ } \hat p_0 = \hat p_{n+1} \equiv 0\text{)}.
\end{equation}
In other words, we assume that the first and last oscillator (i.e.~the
oscillators numbered $1$ and $n$) are attached to a fixed wall.  

The treatment of this Hamiltonian in the WQS approach is very similar to the previous subsection.
In fact, the CCs~\eqref{CCpq1}-\eqref{CCpq2} remain the same (with $\hat{H}_P$ replaced by
$\hat{H}_{FW}$, but with $\hat q_0 = \hat q_{n+1} = 0$).
These different boundary conditions lead to a different type of transform. 
Instead of a discrete Fourier transform, we now use a discrete sine transform.
More explicitly, we introduce the following transformations of the 
(self-adjoint) operators $\hat q_r$ and $\hat p_r$:
\begin{align}\label{q-and-p}
\hat{q}_r &=  \sum_{j=1}^{n} \sqrt{\frac{\hbar}{m(n+1)\tilde\omega_j}}
\sin\bigl( \frac{rj\pi}{n+1}   \bigr)  \big( a_{j}^++a_j^-\big),\\
\hat{p}_r &= i \,
\sum_{j=1}^{n}\sqrt{\frac{m\tilde\omega_j\hbar}{n+1}}
\sin\bigl( \frac{rj\pi}{n+1}   \bigr) 
 \big( a_{j}^+-a_{j}^-\big), 
\end{align}
where the $\tilde\omega_j$ are positive numbers given by 
\begin{equation}
\tilde\omega_j^2 = 
{ \omega^2 + 2c-2c\cos\bigl(\frac{j\pi}{n+1}\bigr)} 
= {\omega^2 + 4c\sin^2\bigl( \frac{j\pi}{2(n+1)}  \bigr)}.
\end{equation}
The operators $a^\pm_j$ satisfy the adjointness conditions
\begin{equation}
\label{a-adj}
(a^\pm_j)^\dagger = a^\mp_j.
\end{equation}  
In terms of these new operators, the Hamiltonian~\eqref{H_FW} is given by:
\begin{equation}
\hat{H}_{FW}=\sum_{j=1}^{n} \frac{\hbar \tilde\omega_j}{2} 
(a_j^- a_j^+ + a_j^+ a_j^-),
\label{H2aa}
\end{equation}
and the compatibility conditions become
\begin{equation}
\Bigl[ \sum_{j=1}^{n} \tilde\omega_j (a_j^- a_j^+ + a_j^+ a_j^-) , a_k^\pm \Bigr]=
\pm 2 \tilde\omega_k a_k^\pm, \quad \ (k=1,2,\ldots,n). 
\label{algrelations2}
\end{equation}

So the algebraic expression of the Hamiltonian, and the CCs, are the same in the two
cases considered, apart from the replacement $\omega_j \rightarrow \tilde\omega_j$. 
This implies that the algebraic solutions will be similar, even though the conclusions
about physical properties will be different due to the different numerical values
of the numbers $\omega_j$ and $\tilde\omega_j$.

\subsection{The $\gl(1|n)$ solution}

In was shown in~\cite{LSV06} that the triple relations~\eqref{algrelations1} involving 
both anti-commutators and commutators have a solution in terms of generators
of the Lie superalgebra $\gl(1|n)$~\cite{Kac1}.
More explicitly, let $\gl(1|n)$ be the Lie superalgebra with standard basis
elements $e_{jk}$ ($j,k=0,1,\ldots,n$) where $e_{k0}$ and $e_{0k}$ ($k=1,\ldots,n$)
are odd elements ($\deg e_{0k} = \deg e_{k0} = 1$) and the remaining basis elements are even (having
degree 0), with bracket
\begin{equation}
\lb e_{ij}, e_{kl} \rb = \delta_{jk} e_{il} - (-1)^{\deg(e_{ij})\deg(e_{kl})} \delta_{il}e_{kj},
\label{eij}
\end{equation}
and star condition $e_{ij}^\dagger=e_{ji}$. 
Then a solution of~(\ref{algrelations1}) is provided by
\begin{equation}
	a_j^- = \sqrt{\frac{2\beta_j}{\omega_j}}\;  e_{j0}, \quad
a_j^+ = \sqrt{\frac{2\beta_j}{\omega_j}}\;  e_{0j} 
\quad(j=1,\ldots,n)
\label{solution1}
\end{equation}
where
\begin{equation}
\beta_j = -\omega_j + \frac{1}{n-1}\sum_{k=1}^n \omega_k,
\quad (j=1,\ldots,n).
\label{beta1}
\end{equation}
All these numbers $\beta_j$ should be nonnegative. By the periodic boundary conditions,
the $\beta_j$'s satisfy $\beta_{n-j}=\beta_j$, and
\begin{equation}
\beta_1 > \beta_2 > \cdots > \beta_{\lfloor n/2\rfloor}, \quad \beta_{\lfloor n/2\rfloor} \leq
\beta_{\lfloor n/2\rfloor+1 }< \cdots < \beta_n.
\end{equation}
It was analysed in~\cite{LSV06} that all these $\beta_j$'s are indeed positive provided 
the coupling constant $c$ lies in a certain interval $[0,c_0[$, with $c_0$ some critical
value depending upon $n$. 

In the case of fixed wall boundary conditions, the analysis is slightly different.
A solution of~(\ref{algrelations2}) is given by
\begin{equation}
	a_j^- = \sqrt{\frac{2\tilde\beta_j}{\tilde\omega_j}}\;  e_{j0}, \quad
a_j^+ = \sqrt{\frac{2\tilde\beta_j}{\tilde\omega_j}}\;  e_{0j} 
\quad(j=1,\ldots,n)
\label{solution2}
\end{equation}
where in this case
\begin{equation}
\tilde\beta_j = -\tilde\omega_j + \frac{1}{n-1}\sum_{k=1}^n \tilde\omega_k,
\quad (j=1,\ldots,n).
\label{beta2}
\end{equation}
Again, all these numbers $\tilde\beta_j$ should be nonnegative.  First, note that
for $c>0$ one has that $\tilde\omega_1 < \tilde\omega_2 < \cdots < \tilde\omega_n$ and hence
that
\begin{equation}\label{beta_fw_ineq}
\tilde\beta_1 > \tilde\beta_2>\cdots>\tilde\beta_n.
\end{equation}
Thus all $\tilde\beta_j$ are positive if and only if $\tilde\beta_n$ is positive.
Secondly, for $c=0$ one has that $\tilde\beta_j = \omega/(n-1)>0$,
and since $\tilde\beta_n$ is a continuous function of $c$ there 
exist positive values of $c$ such that $\tilde\beta_n > 0$. Thus, there exists 
(in general) a critical value $\tilde c_0$ such that each $\tilde\beta_j >0$
for $c < \tilde c_0$ and such that for $c=\tilde c_0$ one 
has that $\tilde\beta_n = 0$.   The same upper bound on the critical value
$\tilde c_0$ applies as on $c_0$ in periodic boundary conditions case
since one can mimic the proof of~\cite[Proposition 2]{LSV06}. 
Since for $n = 2$ we have that $\tilde\beta_2 = \sqrt{\omega^2+c}$ there
are no conditions on $c$ in this case. Also for $n = 3$ there are no 
conditions since one can verify (numerically) that in 
this case $\tilde\beta_3 > 0$. In Table~\ref{tab:crit}, we compare the 
critical values $c_0/\omega^2$ of the periodic boundary conditions case with 
the critical values $\tilde c_0/\omega^2$.  One notices that these 
critical values are interleaved, and that they become more and more similar as 
the number of oscillators $n$ increases.

\begin{table}[htb]
\begin{center}
\begin{tabular}{||r|r|r||r|r|r||}
\hline
$n$ & $c_0/\omega^2$ &  $\tilde c_0/\omega^2$& $n$ & $c_0/\omega^2$ & $\tilde c_0/\omega^2$ \\
\hline
4 & 0.9873724357 & 2.1108888881	& 13 & 0.10546881460  &	0.10521909714	\\
5 & 0.7500000000 & 0.7016444817	& 14 & 0.09256321610 & 	0.09509684206	\\
6 & 0.3457442295 & 0.4138598334	& 15 & 0.08687882025 & 	0.08675785013	\\
7 & 0.2982653656 & 0.2921798279	& 16 & 0.07814800074 & 	0.07976866442	\\
8 & 0.2061705212 & 0.2254893243	& 17 & 0.07388896853 & 	0.07382573538	\\
9 & 0.1851128402 & 0.1835156565	& 18 & 0.06760983697 & 	0.06871018095	\\
10 & 0.1464642846 & 0.1547079900 & 19 & 0.06429500840 & 0.06426020587	\\
11 & 0.1343028683 & 0.1337254495 & 20 & 0.05957194222 & 0.06035363583	\\
12 & 0.1134651313 & 0.1177656002 & 21 & 0.05691629341 & 0.05689649085	\\
\hline
\end{tabular}
\caption{Critical values $c_0/\omega^2$ (periodic boundary conditions)
and $\tilde c_0/\omega^2$ (fixed wall boundary conditions).}
\label{tab:crit}
\end{center}
\end{table}

\subsection{A class of $\gl(1|n)$ representations $V(p)$}

In the case of canonical commutations relations, there is essentially only one 
representation of the system, following from the Heisenberg-Weyl algebra satisfied
by the operators $\hat q_r$ and $\hat p_r$.
In the case of WQS, the properties of the systems described by the Hamiltonians $\hat H_P$ and $\hat H_{FW}$
depend on the $\gl(1|n)$ representation considered.
In principle, any unitary representation of $\gl(1|n)$ can be taken into account~\cite{King}.
A simple class of Fock representations $W(p)$~\cite{Palev2} was already investigated in~\cite{LSV06},
for the case of periodic boundary conditions. This class of Fock representations is 
easy to work with, but also rather restricted as the basis vectors involve
``fermionic'' variables only, see~\cite[(4.1)]{LSV06}.

In this paper, we will consider a richer class of representations, the so-called
ladder representations $V(p)$~\cite{King}. 
These representations were considered as a special case in~\cite{LSV07}.
They are characterized by a positive integer~$p$, and are finite-dimensional unitary representations
atypical of type~2~\cite{LSV07}.
A simple notation for the vectors of $V(p)$ is:
\begin{equation}
w(\theta;s)\equiv w(\theta;s_1,s_2, \ldots,s_n),\qquad
\theta\in\{0,1\},\ s_i\in\{0,1,2,\ldots\},\ \hbox{ and } \theta+s_1+\cdots+s_n= p.
\label{newbasis}
\end{equation}
Thus here the basis vectors involve one ``fermionic'' variable $\theta$
and $n$ ``bosonic'' variables $s_i$.
In this notation the highest weight vector is $w(1;p-1,0,\ldots,0)$.

The action of the $\gl(1|n)$ generators on the basis~(\ref{newbasis}) is given 
by ($1\leq k \leq n$)~\cite{King,LSV07}:
\begin{align}
& e_{00} w(\theta; s) = \theta\ w(\theta; s), \label{e00}\\
& e_{kk} w(\theta; s) = s_k\ w(\theta; s),\label{ekk}\\
& e_{k0} w(\theta; s) =\theta \sqrt{s_k+1}\ w(1-\theta; s_1,\ldots,s_k+1,\ldots,s_n), \label{ek0}\\
& e_{0k} w(\theta; s) =(1-\theta)\sqrt{s_k}\ w(1-\theta; s_1,\ldots,s_k-1,\ldots,s_n). \label{e0k}
\end{align}
{}From these one deduces the action of other elements $e_{kl}$.
The basis $w(\theta;s)$ of $V(p)$ is orthogonal, i.e.\ $\langle w(\theta;s), w(\theta';s') \rangle =
\delta_{\theta,\theta'}\delta_{s,s'}$, and with respect to this inner product the action of the generating elements
satisfies the conjugacy relations $e_{k0}^\dagger=e_{0k}$ and $e_{0k}^\dagger=e_{k0}$.

\setcounter{equation}{0}
\section{On the spectrum of the Hamiltonians in the ladder representation}

In this section, we study the spectrum of the Hamiltonians~\eqref{H_P} and~\eqref{H_FW} 
in the ladder representation $V(p)$ of the $\gl(1|n)$ solution.
Although the basis vectors  
of the representation are in both cases eigenvectors of the Hamiltonian,
the spectrum in the two cases is quite different. 

\subsection{Energy eigenvalues in case of periodic boundary conditions}
For the periodic boundary conditions case, the Hamiltonian~\eqref{H1aa} is given,
using~\eqref{solution1}, by:
\begin{equation*}
\hat H_P = \hbar(\beta\, e_{00} + \sum_{k=1}^n \beta_k\, e_{kk}),
\end{equation*}
with $\beta=\sum_{k=1}^n \beta_k = \sum_{k=1}^n \omega_k$. 
Since the action of each $e_{kk}$ is diagonal in the basis
$w(\theta;s)$, see~\eqref{e00} and~\eqref{ekk}, this implies that each
basis vector $w(\theta;s)$ is an eigenvector of $\hat H_P$, or a stationary state.
Indeed, one finds that
\begin{equation*}
\hat H_P\, w(\theta;s) = \hbar(\beta\theta + \sum_{k=1}^n \beta_k s_k )w(\theta;s)
= \hbar E_{\theta,s}\, w(\theta;s).
\end{equation*}

When $c=0$, one has that $\beta_k = \omega/(n-1)$ and $\beta = \omega n/(n-1)$, so
in this case there are only two eigenvalues namely
\begin{equation*}
E_{0,s} = \frac{\omega p}{n-1}\quad\text{and}\quad
E_{1,s} = \frac{\omega p}{n-1} + \omega
\end{equation*}
with multiplicities
\begin{equation*}
\binom{p+n-1}{n-1}\quad\text{and}\quad
\binom{p+n-2}{n-1}
\end{equation*}
respectively.  

When $c>0$, these two levels each split into a number of energy 
levels with lower degeneracies.  Recall that $\beta_k = \beta_{n-k}$. 
So, when $n = 2r$ is even, one can rewrite $E_{\theta,s}$ as 
follows:
\begin{equation}\label{E-rewrite}
E_{\theta,s} = \beta\theta + \sum_{k=1}^{r-1}\beta_k(s_k + s_{n-k}) +
\beta_{r}s_{r} + \beta_n s_n.
\end{equation}
It is then clear that any basis vector $w(\theta;s')$ for which
$s'_k+s'_{n-k} = s_k+s_{n-k}$ ($k=1,\ldots,n-1$) yields the same $\hat H_P$
eigenvalue, independent of the value of $c$, 
i.e.~for such $\theta$ and $s'$ one has that $E_{\theta,s} = E_{\theta,s'}$.
Alternatively, one can say that for a fixed value of $\theta$, $E_{\theta,s}$
is completely determined by
\begin{equation}\label{det-en}
(s_1+s_{n-1},s_2+s_{n-2},\ldots,s_{r-1}+s_{n-r+1},s_{r},s_n),
\end{equation}
which is a composition of $p-\theta$ into $r+1$ parts.  In the same 
way one sees that when $n=2r+1$ is odd, $E_{\theta,s}$ with $\theta$ fixed,
is determined by
\begin{equation}\label{det-en2}
(s_1+s_{n-1},s_2+s_{n-2},\ldots,s_{r-1}+s_{n-r+1},s_{r} + s_{r+1},s_n),
\end{equation}
For 
$0<c\leq c_0$ and $n=2r$ or $n=2r+1$,  
the number of different energy levels is thus in general given by
\begin{equation}
\label{nrel}
\binom{p+r}{p}+\binom{p+r-1}{p-1}.
\end{equation}

One can also say something about the degeneracy of an individual energy level
$\hbar E_{\theta,s}$.  The number of 
compositions of an integer $N$ in two parts is $N+1$.  
Using this fact and~\eqref{det-en} or~\eqref{det-en2} one sees that the degeneracy of 
$\hbar E_{\theta,s}$ is at least:
\begin{equation}\label{en-deg}
\prod_{k=1}^{\lfloor (n-1)/2\rfloor} (s_k + s_{n-k}+1).
\end{equation}

\begin{figure}[htbp] 
	\begin{center}
 \begin{tabular}{ccc}
 (a) $n=4$ &\vspace{12mm} & (b) $n=5$ \\
 \includegraphics[width=6cm]{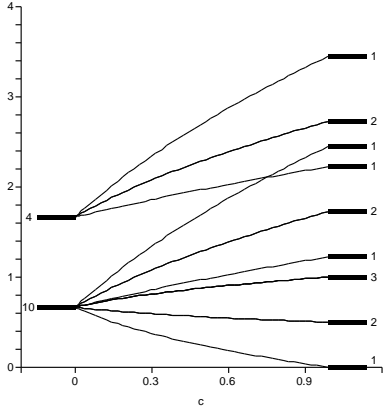} &&
 \includegraphics[width=6cm]{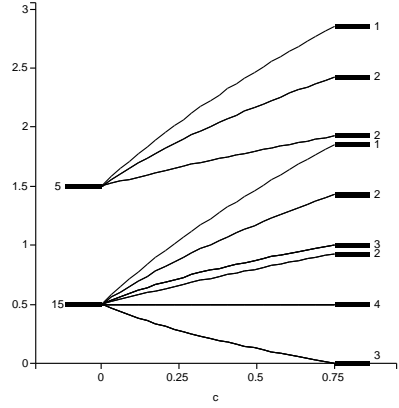}
 \end{tabular}
	\caption{(a) 
	The energy levels of the quantum system with periodic boundary conditions 
	for $n=4$ in the representation $V(p)$ with $p=2$ and 
	$\hbar=\omega=1$; $c$ ranges from 0 to $c_0$.  
	The vertical axis gives the energy values and 
	the numbers next to the levels refer to the multiplicity.  When $c=0$
	there are only two energy levels with multiplicities 10 and 4.  When
	$0<c<c_0$ there are (in general) 9 energy levels with multiplicity
	1, 2 or 3.  (b)  The same illustration for $n=5$.  Note that since 
	$\lfloor 4/2\rfloor = \lfloor 5/2\rfloor$ the number of different 
	energy levels is (in general) unchanged for $0<c<c_0$.}
	\label{fig:H}
\end{center}
\end{figure}

Figure~\ref{fig:H} shows two spectra of $\hat H_P$ for $n=4$ and $n=5$
with $p=2$ in both cases.  These figures confirm our findings: in general
for $0<c\leq c_0$ there are 
\begin{equation*}
\binom{2+2}{2}+\binom{2+2-1}{1} = 6 + 3 = 9 
\end{equation*}
different energy levels.  For $n=4$, the energy level $\hbar E_{\theta,s}$ with 
threefold degeneracy is given by
\begin{equation*}
E_{0;2,0,0,0} = E_{0;0,0,2,0} = E_{0;1,0,1,0},
\end{equation*}
whereas the fourfold degenerate energy level for $n=5$ is determined by
\begin{equation*}
E_{0;1,1,0,0,0} = E_{0;1,0,1,0,0} = E_{0;0,1,0,1,0} = E_{0;0,0,1,1,0}.
\end{equation*}

However, as can be seen from the figure, for particular values of $c$ 
it may happen that the multiplicity of some eigenvalues is greater than
stated in~\eqref{en-deg} (and hence the number of different 
energy levels is smaller than~\eqref{nrel}).   It is infeasible to
obtain  analytical expressions for the particular values of $c$ for which this will happen.

Now, we briefly turn our attention to the minimal and maximal energy eigenvalues.
For $c>0$, the values $\beta_k$ satisfy the following inequalities:
\begin{equation*}
\beta_1 > \beta_2 >\cdots > \beta_{r},\quad
\beta_{r} \leq \beta_{r+1} < \cdots < \beta_n\quad\text{and}\quad \beta_1 < \beta_n,
\end{equation*}
and equality between $\beta_{r}$ and $\beta_{r+1}$ only occurs when $n$ is odd.
{}From this, it immediately follows that the maximal eigenvalue of 
the Hamiltonian in the ladder representation $V(p)$ is given by $\hbar(\beta+(p-1)\beta_n)$,
and the corresponding eigenvector is $w(\theta,s)$ with $\theta = 1$ and $s_j = \delta_{j,n}(p-1)$.
It is also immediately clear that this eigenvalue is nondegenerate.  The minimum 
eigenvalue is given by $\hbar p \beta_{r}$.  An eigenvector is given by $w(\theta;s)$
with $\theta = 0$ and $s_j = \delta_{j,r} p$. When $n$ is even this eigenvalue 
is nondegenerate, but when $n$ is odd it is $(p+1)$-fold degenerate, since $p+1$
is the number of compositions of $p$ into two parts. 
Note that the minimal eigenvalue approaches 0 as $c$ tends to $c_0$.
This is also illustrated by
Figure~\ref{fig:H}.

\subsection{Energy eigenvalues in case of fixed wall boundary conditions}

\begin{figure}[htbp]
	\begin{center}
 \begin{tabular}{ccc}
 (a) $n=4$ &\vspace{12mm} & (b) $n=5$ \\
 \includegraphics[width=6cm]{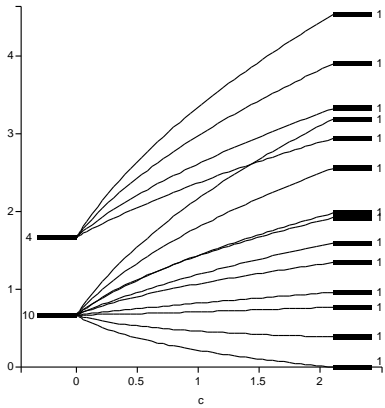} &&
 \includegraphics[width=6cm]{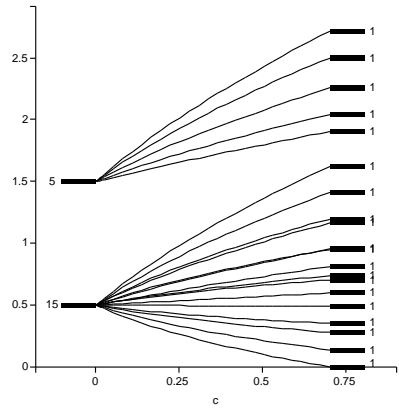}
 \end{tabular}
	\caption{(a) 
	The energy levels of the quantum system with fixed wall boundary conditions for $n=4$ in the representation $V(p)$ with $p=2$ and 
	$\hbar=\omega=1$; $c$ ranges from 0 to $\tilde c_0$.  
	The vertical axis gives the energy values and 
	the numbers next to the levels refer to the multiplicity.  When $c=0$
	one recovers the results of the periodic boundary case.  When
	$0<c<\tilde c_0$  however, there are (in general) 14 energy levels each with multiplicity
	1.  (b)  The same illustration for $n=5$, but now there are 20 nondegenerate energy levels.  }
	\label{fig:H-fw}
\end{center}
\end{figure}

In the case with fixed wall boundary conditions the Hamiltonian~\eqref{H2aa} becomes,
using~\eqref{solution2},
\begin{equation*}
\hat H_{FW} = \hbar(\tilde\beta\, e_{00} + \sum_{k=1}^n \tilde\beta_k\, e_{kk}),
\end{equation*}
with $\tilde\beta = \sum_{k=1}^n \tilde\beta_k$;
so clearly one has that the basis vectors of the representation are eigenvectors:
\begin{equation*}
\hat H_{FW}\, w(\theta;s) = \hbar(\tilde\beta\theta + \sum_{k=1}^n 
\tilde\beta_k s_k )w(\theta;s)
= \hbar \tilde E_{\theta,s}\, w(\theta;s).
\end{equation*}
The analysis of these eigenvalues turns out to be easier than in the previous case,
because the $\tilde\beta_j$'s do not satisfy any symmetry relations, but
only the inequalities~\eqref{beta_fw_ineq}.
For $c=0$ one recovers the same two energy values and degeneracies as 
before, but in general for $c>0$, the energy levels are nondegenerate
since the $\tilde\beta_k$ do not show any symmetry.  
Figure~\ref{fig:H-fw} illustrates this fact.

In this case, the maximal eigenvalue of $\hat H_{FW}$
is $\hbar(\tilde\beta + (p-1)\tilde\beta_1)$, while the minimal eigenvalue is
given by $\hbar\, p\, \tilde\beta_n$.  This follows immediately from the 
inequalities~\eqref{beta_fw_ineq}.  Of course, these eigenvalues are {always} 
nondegenerate.

\setcounter{equation}{0}
\section{On the eigenvalues and eigenvectors of the position operators}

We now turn to the study of the spectrum of the position operators $\hat q_r$,
which are, under the solutions~\eqref{solution1} or~\eqref{solution2},  
quite arbitrary odd elements of the Lie 
superalgebra $\gl(1|n)$.  This is precisely the topic of~\cite{LSV07} where
the eigenvalues (and eigenvectors) of the position operators $\hat q_r$ were determined
for arbitrary unitary irreducible representations of the Lie superalgebra $\gl(1|n)$.
Thus the method developed there will be applied both to the case of periodic and 
fixed wall boundary conditions.

\subsection{Position eigenvalues in the case of periodic boundary conditions}

The position operator $\hat q_r$, given by~\eqref{qr1}, can be written as
\begin{align}
\hat q_r  & = \sqrt{\frac{\hbar}{m n}} \sum_{j=1}^n \left( 
\gamma_j\, e^{2\pi i jr/n} e_{j0} + \gamma_j\, e^{-2\pi i jr/n} e_{0j}
 \right) \label{q-e2} \\
& = \sqrt{\frac{\hbar \gamma}{m n}} (E_{n0}^{(r)} + E_{0n}^{(r)}), \label{q_r-E}
\end{align}
where we use the notation
\begin{equation}
\gamma_j = \sqrt{\beta_j}/\omega_j \quad (j=1,\ldots,n) \hbox{ and }
\gamma = \gamma_1^2 + \cdots + \gamma_n^2.
\end{equation}
The odd $\gl(1|n)$ operators
\begin{equation}
\label{En0}
E_{n0}^{(r)} = \frac{1}{\sqrt{\gamma}} \sum_{j=1}^n \gamma_j\, e^{2\pi i j r/n}e_{j0} 
= \sum_{j=1}^n U_{nj}^{(r)} e_{j0},
\qquad
E_{0n}^{(r)} = \sum_{j=1}^n {U_{nj}^{(r)}}^* e_{0j}
\end{equation}
are part of a more general set of operators
\begin{equation}
\label{def-E}
E_{j0}^{(r)}  = \sum_{l=1}^n U_{jl}^{(r)}\, e_{l0}\ \text{ and }\
E_{0j}^{(r)}  = \sum_{l=1}^n {U_{jl}^{(r)}}^*\, e_{0l}\qquad (1\leq j \leq n),
\end{equation}
where $U = (U_{jl}^{(r)})_{1\leq j\leq n,1\leq l\leq n}$ is a unitary $n\times n$ matrix
determined by the coefficients in~\eqref{En0} and in
\begin{equation}
\label{Ej0}
E_{j0}^{(r)}  = \frac{1}{ \sqrt{\frac{1}{\gamma_1^2+\cdots+ \gamma_{j}^2} 
+ \frac{1}{\gamma_{j+1}^2}} } 
\left( \sum_{l=1}^{j} 
\frac{e^{2\pi i r l/n}}
     {\gamma_1^2+\cdots + \gamma_{j}^2} \gamma_l e_{l0} 
     - \frac{1}{\gamma_{j+1}} e^{2\pi i r (j+1)/n} e_{j+1,0}\right)
\end{equation}
for $j=1,2,\ldots,n-1$.
The new operators $E_{j0}^{(r)}$ and $E_{0j}^{(r)}$ satisfy the same defining relations 
as the elements $e_{j0}$ and $e_{0j}$; in other words, they generate the Lie superalgebra $\ssl(1|n)$~\cite{LSV07}.
Thanks to the introduction of these new operators, one can determine the eigenvalues
and eigenvectors of the position operators $\hat q_r$. 
This follows from the observation that~\eqref{q_r-E} is essentially an element of $\gl(1|1)$
in the decomposition $\gl(1|n) \rightarrow \gl(1|1) \oplus \gl(n-1)$. 
The following result was obtained in~\cite{LSV07}:
\begin{prop}
In the representation $V(p)$, all operators $\hat q_r$ ($r=1,2,\ldots,n$) have
the same spectrum. The operator $\q_r$ has $2p+1$ distinct eigenvalues 
given by 
\begin{equation}
\pm x_K=\pm \sqrt{\frac{\hbar\gamma}{m n}(p-K)}, \qquad K=0,1,\ldots,p.
\end{equation} 
The multiplicity of the eigenvalue $\pm x_K$ is $\binom{n-2+K}{K}$.  
\end{prop}
Let us also briefly describe the corresponding orthonormal eigenvectors 
$\psi_{r,\pm x_K, t}$ (with $t$ a multiplicity label) corresponding to
the eigenvalue $\pm x_K$ (see~\cite{LSV07} for the details). For $K \neq p$, one has:
\begin{equation}
\label{Vpsi}
\psi_{r,\pm x_K,t} = \frac{1}{\sqrt{2}} v(1;t_1,\ldots,t_{n-1},p-1-K)
\pm \frac{1}{\sqrt{2}} v(0;t_1,\ldots,t_{n-1},p-K),
\end{equation}
where $t_1+\cdots+t_{n-1}=K$. 
For the eigenvalue~0 ($K=p$), the eigenvectors read
\begin{equation}
\label{Vpsi0}
\psi_{r,0,t}= v(0; t_1,\ldots,t_{n-1},0),\qquad t_1+\cdots+t_{n-1}=p.
\end{equation}
We still need to describe the vectors $v(\theta;t)$ of $V(p)$ in terms
of the basis vectors $w(\theta;s)$. 
Essentially, the vectors $v(\theta;t)$ are chosen in such a way that the
action of the ``new'' $\gl(1|n)$ elements $E_{jk}^{(r)}$ on $v(\theta;t)$ are the same
as the action of the ``old'' $\gl(1|n)$ elements $e_{jk}$ on $w(\theta;s)$.
For the new highest weight vector, one has:
\begin{align}
v(1;p-1,0,\ldots,0)& =\frac{1}{(\gamma_1^2+\gamma_2^2)^{(p-1)/2}}
\sum_{u=0}^{p-1} (-1)^{u} e^{-2\pi i r u/n} \sqrt{\binom{p-1}{u}}\nn\\
&\qquad \times \gamma_1^{p-1-u}\gamma_2^{u}\; w(1;u,p-1-u,0,\ldots,0).
\label{V-LambdaE}
\end{align}
The remaining vectors $v(\theta;t)$ are given by
\begin{align}
v(\phi;t_1,\ldots,t_n) & = \frac{1}{\sqrt{N}}
(E_{n,n-1}^{(r)})^{p-\phi-\sum_{j=1}^{n-1}t_j} 
(E_{n-1,n-2}^{(r)})^{p-\phi-\sum_{j=1}^{n-2}t_j} 
\cdots  \nn\\
& \cdots  
(E_{32}^{(r)})^{p-\phi-\sum_{j=1}^2 t_j} 
(E_{21}^{(r)})^{p-\phi-\sum_{j=1}^1 t_j} 
(E_{10}^{(r)})^{1-\phi} 
v(1;p-1,0,\ldots,0); \label{Vv-phi} \\
N &= p^{1-\phi} \prod_{k=1}^{n-1} (p-\phi - \sum_{j=1}^k t_j)! (t_k+1)_{p-\phi - \sum_{j=1}^k t_j},
\end{align}
with $(a)_n = a(a+1)\cdots(a+n-1)$ the rising factorial symbol.

Some specific properties of the spectrum of $\hat q_r$ will be considered together with
those for the fixed wall boundary conditions, in the next subsection.
The expressions for the eigenvectors will be used in the following section,
where position probability distributions are studied.

\subsection{Position eigenvalues in the case of fixed wall boundary conditions}

The formal part of the analysis will be rather similar to that of the
previous case of periodic boundary conditions. But the outcome will show one major difference:
the spectrum of $\hat q_r$ is now dependent (albeit
in a simple way) on the position $r$ of the oscillator in the chain.  This
result is completely in accordance with physical intuition as the oscillators
in the chain are clearly no longer equivalent, since we have two distinguished
(and equivalent) oscillators which mark the beginning and the end of the chain.

Under the solution~\eqref{solution2} of~\eqref{algrelations2}, 
the position operator $\hat q_r$, which is given by~\eqref{q-and-p}, becomes
\begin{align}
\label{q_ej0}
\hat{q}_r & =  \sqrt{\frac{2\hbar}{m(n+1)}}\sum_{j=1}^{n} 
\Bigl(\sin\bigl( \frac{rj\pi}{n+1}\bigr) \tilde\gamma_j e_{j0} +
\sin\bigl( \frac{rj\pi}{n+1}\bigr) \tilde\gamma_j e_{0j}\Bigr) \\ \label{q_ej02}
& = \sqrt{\frac{2\hbar N_r^2}{m(n+1)}}(\tilde E_{n0}^{(r)} + \tilde E_{0n}^{(r)}).
\end{align}
Here, we used the abbreviations
\begin{equation}
\tilde\gamma_j = \frac{\sqrt{\tilde\beta_j}}{\tilde\omega_j} \hbox{ and }
N_r^2 =  \sum_{j=1}^{n} 
\sin^2\bigl( \frac{rj\pi}{n+1}\bigr) \tilde\gamma_j^2 .
\end{equation}
Comparing with~\eqref{q_r-E}, it is clear that $\tilde E_{n0}^{(r)}$ is obtained from 
$E_{n0}^{(n)}$ by performing the substitutions
\begin{equation}
\label{subs}
\gamma_j \leadsto \tilde\gamma_j \sin(\frac{r j \pi}{n+1})\quad (j=1,\ldots,n).
\end{equation}
The same substitution can be used for $E_{j0}^{(n)} \rightarrow \tilde E_{j0}^{(r)}$ and
$E_{0j}^{(n)} \rightarrow \tilde E_{0j}^{(r)}$.
Then the analysis of the eigenvalues and eigenvectors of $\hat q_r$ is determined
by the same technique as in the previous subsection.
In particular:
\begin{prop}
In the representation $V(p)$, the operators $\hat q_r$ ($r=1,2,\ldots,n$) have
a spectrum depending upon $r$. The operator $\q_r$ has $2p+1$ distinct eigenvalues 
given by 
\begin{equation}
\pm x_K=\pm \sqrt{\frac{2\hbar N_r^2}{m (n+1)}(p-K)}, \qquad K=0,1,\ldots,p.
\label{spec-qr}
\end{equation} 
The multiplicity of the eigenvalue $\pm x_K$ is $\binom{n-2+K}{K}$.  
\end{prop}

The corresponding orthonormal eigenvectors $\psi_{r,\pm x_K,t}$ are given by
the same expressions~\eqref{Vpsi}-\eqref{Vpsi0}, where one should use the 
replacement~\eqref{subs} in~\eqref{V-LambdaE} 
(with $r=n$ so that the complex exponential reduces to 1) and 
$E_{jk}^{(n)} \rightarrow \tilde E_{jk}^{(r)}$ in~\eqref{Vv-phi}.

Although the spectrum of $\hat q_r$ is now dependent on the position of the
oscillator in the chain (i.e.~on $r$) through the constant $N_r$
in~\eqref{spec-qr}, one still has the expected symmetry that the spectra of
$\hat q_r$ and $\hat q_{n-r+1}$ coincide, since
$N_r^2 = N_{n-r+1}^2$. 

\begin{figure}
\begin{center}
\includegraphics[width=6cm]{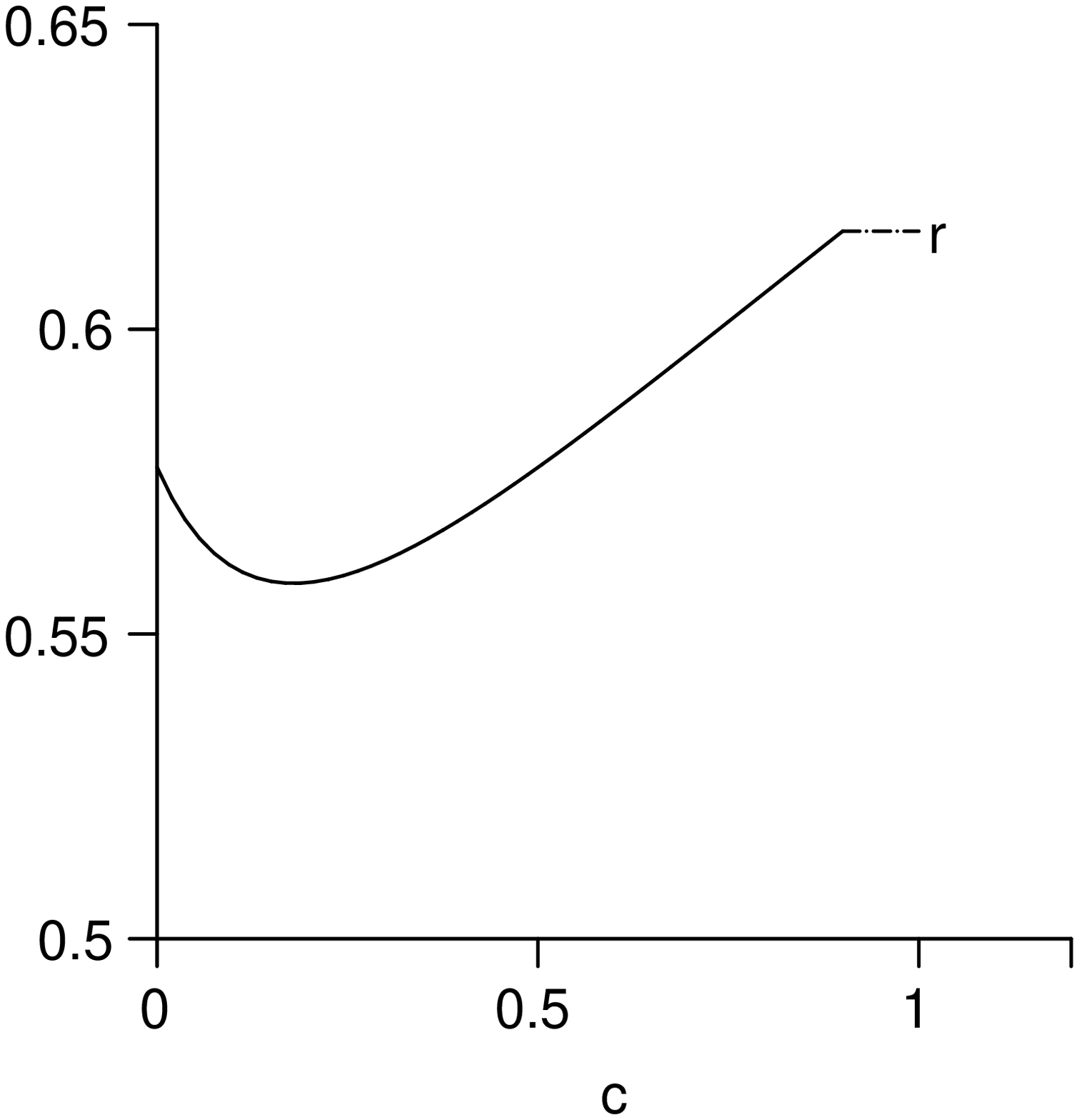}
\includegraphics[width=6cm]{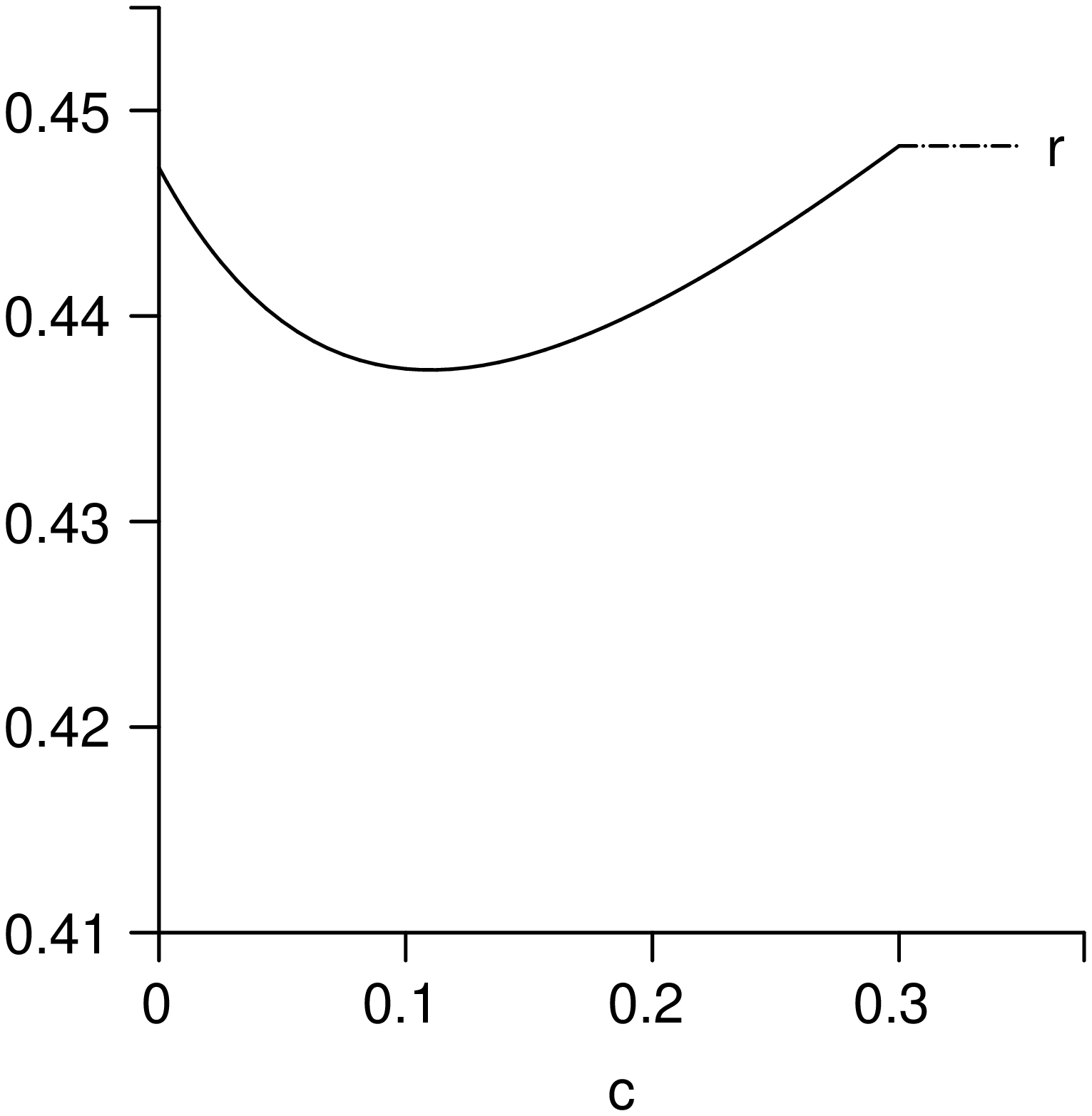}
\includegraphics[width=6cm]{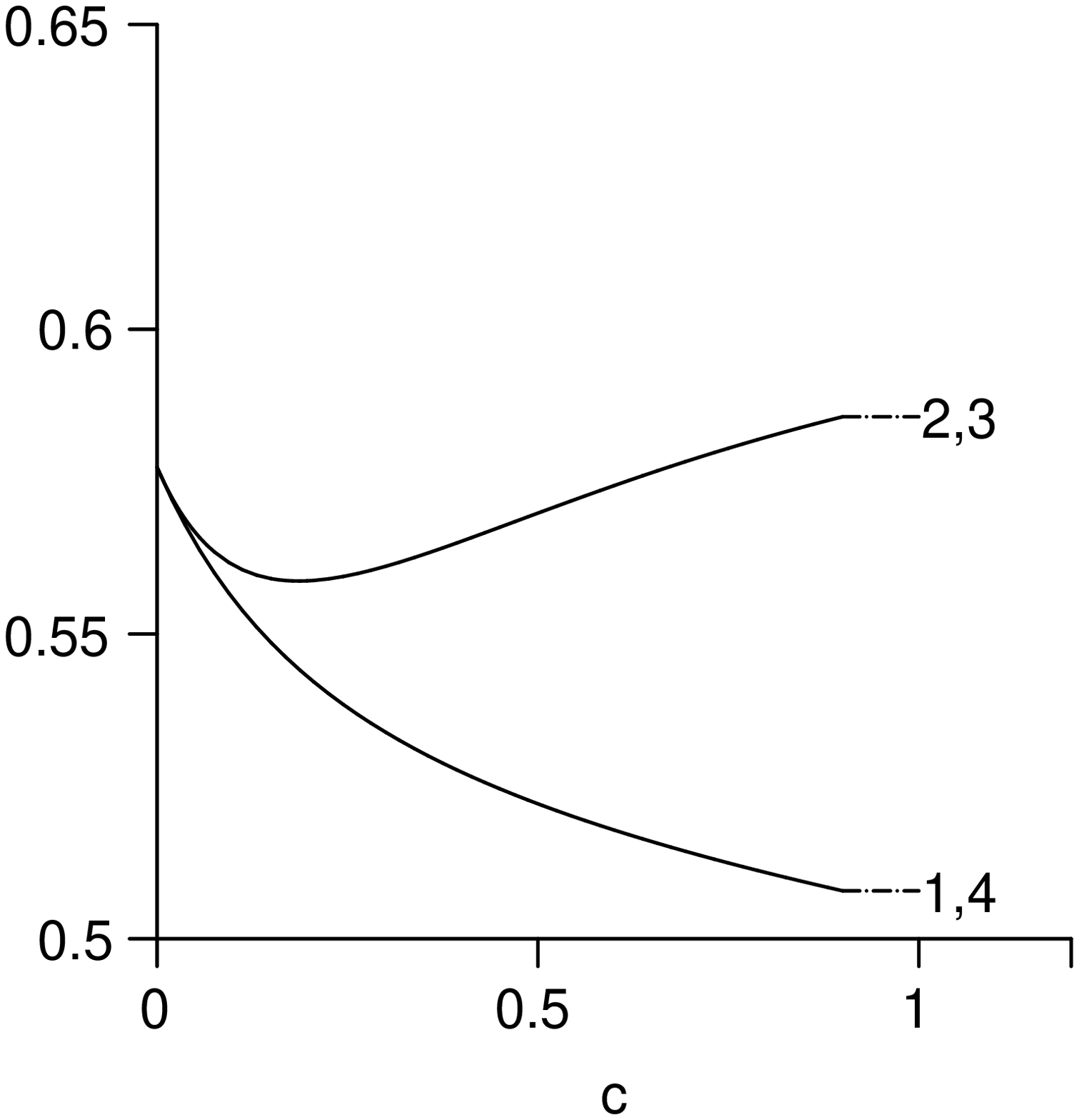}
\includegraphics[width=6cm]{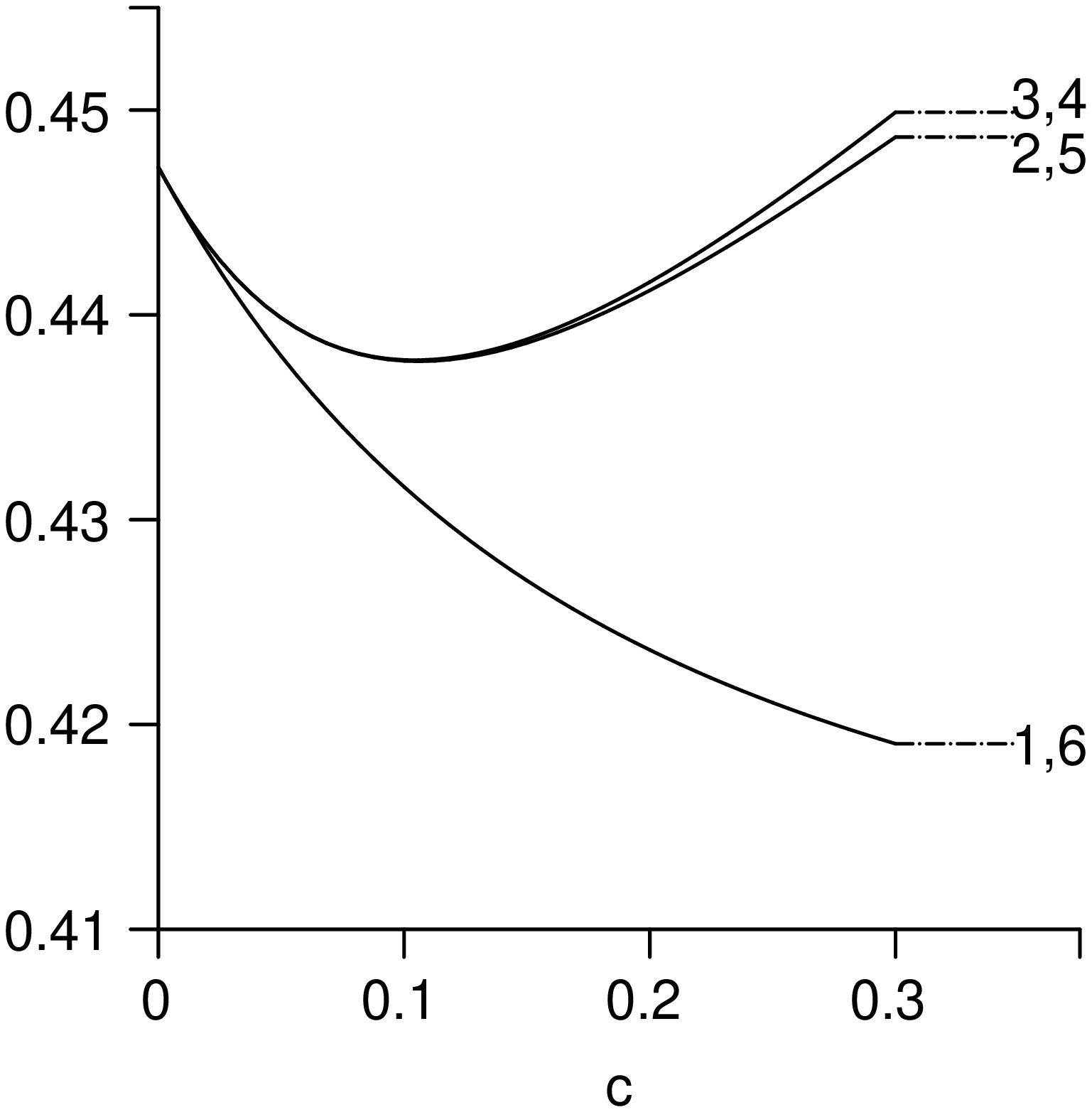}
\caption{Indication of the spectrum of various position operators 
for the two chains, as a
function of the coupling constant $c$ and 
with $m=\hbar=\omega = 1$.
The horizontal axis in these figures represents $c$. 
In the left column $n=4$, while in the right column $n=6$.
In the top row $\sqrt{\gamma/n}$, i.e.~the eigenvalue for an arbitray position 
operator ($r=1,\ldots,n)$
in a chain with periodic boundary conditions
with $K = p-1$, is shown.
The figures in the second row show $\sqrt{\frac{2}{n+1}}N_r$, with
$r$ indicated next to the graph. In other words, they give the 
eigenvalue of the position operators 
 in a chain with fixed wall boundary conditions (again with $K=p-1$).
}
\label{fig:spec-q}
\end{center}
\end{figure}

The dependence on $r$ of the spectrum of $\hat q_r$ is completely determined by $N_r$,
which on its turn depends on the coupling constant~$c$.
In Figure~\ref{fig:spec-q}, we will plot some of these eigenvalues as a function of~$c$.
For these plots, we use $m=\hbar=\omega=1$, and choose $K=p-1$ (then the eigenvalue
expression is independent of $p$).
We plot the cases of periodic boundary conditions 
and of fixed wall boundary conditions  in two different figures.
Thus for periodic boundary conditions, we plot the value $\sqrt{\frac{\gamma}{{n}}}$ as
a function of $c$; for fixed wall boundary conditions, we plot the values 
$\sqrt{\frac{2}{n+1}}N_r$ as a function of $c$.

This figure, together with other numerical experiments, suggests the following
about the range of the spectrum, independent of any measurement probabilities:
\begin{itemize}
\item When the coupling constant $c$ is fixed one has that 
	$N_1 \leq N_2 \leq \cdots \leq N_{\lceil \frac{n}{2} \rceil}$.  This means that the
	spectrum of oscillators close to the wall is more centered around their equilibrium 
	positions than for oscillators in the middle of the chain.  
	This is an intuitively clear
	result as oscillators in the middle of the chain do not \lq\lq feel\rq\rq\ the 
	walls as much as oscillators close to the walls do.
\item  Viewed as a function of $c$, $N_1$ is decreasing.  
	When the coupling constant $c$ increases, it is clear
	that the movements of the first oscillator will become more restricted, 
	resulting in a spectrum closer to its equilibrium position.
\item The spectra of position operators associated with oscillators in the middle 
	of the chain seem to become very similar to one another (as $n$ increases). 
	This is in line with intuition as these oscillators are more or less equivalent
	with respect to their distance to the wall.  Related to this fact one sees that the
	spectrum of an arbitrary position operator associated with an oscillator 
	from a chain with periodic boundary conditions is also very similar to the spectrum
	of an oscillator in the middle of the chain with fixed wall boundary conditions.
\item  For oscillators other than the first and the last, at first the spectrum moves closer to
	the equilibrium positions, but then seems to widen again.  
	This effect could be related to an increasing effect of
	``collective'' motions (lattice vibrations) as
	the coupling constant becomes larger.
\end{itemize}

In the above figures, we have concentrated on a particular $K$-value, i.e.\ on a particular
eigenvalue. Of course, the complete spectrum of each $\hat q_r$ operator is easy to describe:
it is simply a constant times $\pm\sqrt{p-K}$, $K=0,1,\ldots,p$. This is a simple distribution
that will be depicted in the figures of the following section.

\setcounter{equation}{0}
\section{Position probability distributions in the ladder representations}

In this section, we shall establish some facts about the position probability
distributions of the two systems.  It is a well known fact (postulate) of quantum mechanics
that when measuring an observable, the measurement always yields an eigenvalue 
of the (self-adjoint) operator associated with that observable.  The probability of
measuring a certain eigenvalue when the system is in a certain state is determined by
the expansion of that state in terms of (orthonormal) eigenvectors of the operator at hand.

We now assume that our system is in a stationary state $w(\theta;s)$ and
we wish to determine the probabilities of obtaining the different eigenvalues 
$x_K$ of an operator $\hat q_r$.  In the ladder representation $V(p)$, the 
operator $\hat q_r$ has $2p+1$ distinct eigenvalues 
$\pm x_K=\pm \sqrt{A_r(p-K)}$, where  $0 \leq K \leq p$ and with
$A_r$ a constant that in the case of periodic boundary conditions is
independent of $r$, but does depend on $r$ in the case of fixed wall 
boundary conditions. 
The multiplicity of the eigenvalue $\pm x_K$ is $\binom{n-2+K}{K}$, and 
the orthonormal eigenvectors $\psi_{r,\pm x_K, t}$ (with $t$ a multiplicity label) 
have been given in the previous section.
Each eigenvector $\psi_{r,\pm x_K,t}$ can be expanded in terms of the stationary 
states $w(\theta;s)$:
\begin{equation*}
\psi_{r,\pm x_K,t} = \sum_{\theta,s} C^{\theta,s}_{r,\pm x_K,t} w(\theta;s).
\end{equation*}
Using (for the case of periodic boundary conditions) equations~\eqref{V-LambdaE} 
and~\eqref{Vv-phi} the coefficients $C^{\theta,s}_{r,\pm x_K,t}$
in this expansion can be computed explicitly (the same holds of course
for the case of fixed wall boundary conditions).  Using orthonormality of 
the eigenvectors and the basis vectors, we immediately have that 
\begin{equation*}
w(\theta;s) = \sum_K \sum_{t_1+\cdots+t_{n-1} = K} (C^{\theta,s}_{r,\pm x_K,t})^* \psi_{r,\pm x_K,t}.
\end{equation*}
When the quantum system is in the fixed stationary state $w(\theta,s)$, the 
probability of measuring for $\hat q_r$ the eigenvalue $\pm x_K$ 
is given by:
\begin{equation}
\label{pos-prob}
P(\theta,s,r,\pm x_K) = \sum_{t_1+\cdots+t_{n-1} = K} \left| C^{\theta,s}_{r,\pm x_K,t}\right|^2.
\end{equation}
{}From~\eqref{Vpsi}  it is immediately 
clear that 
\begin{equation}
P(\theta,s,r, x_K) =P(\theta,s,r, -x_K). 
\end{equation}

When $p=1$, one can determine explicit expressions for the vectors $v(\phi;t)$.
In this case when $\phi=0$ and there is exactly 
one $t_j = 1$, we introduce the shorthand notation $v(0;1^j)$ for 
this vector.  In the case of periodic boundary conditions, the position 
probabilities do not depend on the position $r$ of the oscillator in 
the chain, so we work with $r=n$ so that all complex exponentials 
occurring in the various expressions reduce to 1.  In this case, the highest 
weight vector is given by
\begin{equation*}
v(1;0,\ldots,0) = w(1;0,\ldots,0),
\end{equation*}
and using induction on $k$, (since $v(0;1^{k+1}) = E_{k+1,k}^{(n)} v(0;1^k)$) one can prove that
\begin{equation*}
\begin{split}
v(0;1^k) & = \frac{1}{\sqrt{\gamma_1^2+\cdots+\gamma_{k+1}^2}
}
\bigl(\sum_{j=1}^k \frac{\gamma_j\gamma_{k+1}}{\sqrt{\gamma_1^2+\cdots+\gamma_{k}^2}} w(0;1^j) - 
\sqrt{\gamma_1^2+\cdots+\gamma_{k}^2}\, w(0;1^{k+1})\bigr),\quad\text{when}\  k \neq n \\
v(0;1^n) & = \frac{1}{\sqrt{\gamma_1^2+\cdots+\gamma_n^2}} \sum_{j=1}^n \gamma_j w(0;1^j).
\end{split}
\end{equation*}
We can now immediately give the expansion of the eigenvectors of $\hat q_n$ 
in terms of the basis vectors $w(\theta;s)$:
\begin{equation*}
\begin{split}
\psi_{n,\pm x_0,t} & = \frac{1}{\sqrt{2}} w(1;0,\ldots,0) 
\pm 
\frac{1}{\sqrt{2\gamma}} \sum_{j=1}^n \gamma_j w(0;1^j),\quad t = (\overbrace{0,\ldots,0}^{n-1}). \\
\psi_{n,x_1,t} &  =  \frac{1}{\sqrt{\gamma_1^2+\cdots+\gamma_{k+1}^2}
}\bigl(\sum_{j=1}^k \frac{\gamma_j\gamma_{k+1}}{\sqrt{\gamma_1^2+\cdots+\gamma_{k}^2}} w(0;1^j) - 
\sqrt{\gamma_1^2+\cdots+\gamma_{k}^2}\, w(0;1^{k+1})\bigr),\quad t = 1^k,
\end{split}
\end{equation*}
with $1\leq k\leq n-1$. {}From this and~\eqref{pos-prob} it immediately follows that
\begin{equation*}
P(\theta,s,r,x_0) = 
\begin{cases}
\frac{1}{2}\frac{\gamma_k^2}{\gamma} &\text{when}\ \theta=0, s_k = 1 \\
\frac{1}{2} & \text{when}\ \theta = 1,
\end{cases}
\end{equation*}
and hence
\begin{equation*}
P(\theta,s,r,x_1) = 
\begin{cases}
\frac{\gamma-\gamma_k^2}{\gamma} &\text{when}\ \theta=0, s_k = 1 \\
0 & \text{when}\ \theta = 1.
\end{cases}
\end{equation*}

Also for $p=2$ we have been able to determine the position probabilities.
They are as follows:
\begin{equation*}
P(\theta,s,r,x_0) = 
\begin{cases}
\frac{1}{2}\frac{\gamma_k^4}{\gamma^2} &\text{when}\ \theta=0, s_k = 2 \\
\frac{\gamma_k^2\gamma_l^2}{\gamma^2} & \text{when}\ \theta = 0, s_k=s_l = 1\\
\frac{1}{2}\frac{\gamma_k^2}{\gamma} &\text{when}\ \theta = 1, s_k = 1,
\end{cases}
\end{equation*}
\begin{equation*}
P(\theta,s,r,x_1) = 
\begin{cases}
\frac{\gamma_k^2(\gamma-\gamma_k^2)}{\gamma^2}
	&\text{when}\ \theta=0, s_k = 2 \\
\frac{1}{2}\frac{ (\gamma-\gamma_k^2-\gamma_l^2)(\gamma_k^2+\gamma_l^2) + (\gamma_k^2-\gamma_l^2)^2}{\gamma^2} & \text{when}\ \theta = 0, s_k=s_l = 1\\
\frac{1}{2}\frac{\gamma-\gamma_k^2}{\gamma} &\text{when}\ \theta = 1, s_k = 1,
\end{cases}
\end{equation*}
and
\begin{equation*}
P(\theta,s,r,x_2) = 
\begin{cases}
\frac{(\gamma-\gamma_k^2)^2}{\gamma^2} &\text{when}\ \theta=0, s_k = 2 \\
\frac{ (\gamma-\gamma_k^2-\gamma_l^2)\gamma + 2\gamma_k^2\gamma_l^2}{\gamma^2} & \text{when}\ \theta = 0, s_k=s_l = 1\\
0 &\text{when}\ \theta = 1, s_k = 1.
\end{cases}
\end{equation*}

In the case of fixed wall boundary conditions, these position probabilities are
given by the same expressions, subject to the substitutions~\eqref{subs}. 
So, as was to be expected, the position probabilities do depend on the position 
of the oscillator in the chain.

For $p=1$ and $p=2$ it was possible to compute these position probabilities analytically,
but for $p>2$ this becomes infeasible. 
On the other hand, since all coefficients $C^{\theta,s}_{r,\pm x_K,t}$ are known
explicitly (in all cases and for any~$p$), we can numerically compute all position
probabilities. 
We will now examine the plots of some of these position probability distributions.

\begin{figure}
\begin{center}
\includegraphics[width=18cm]{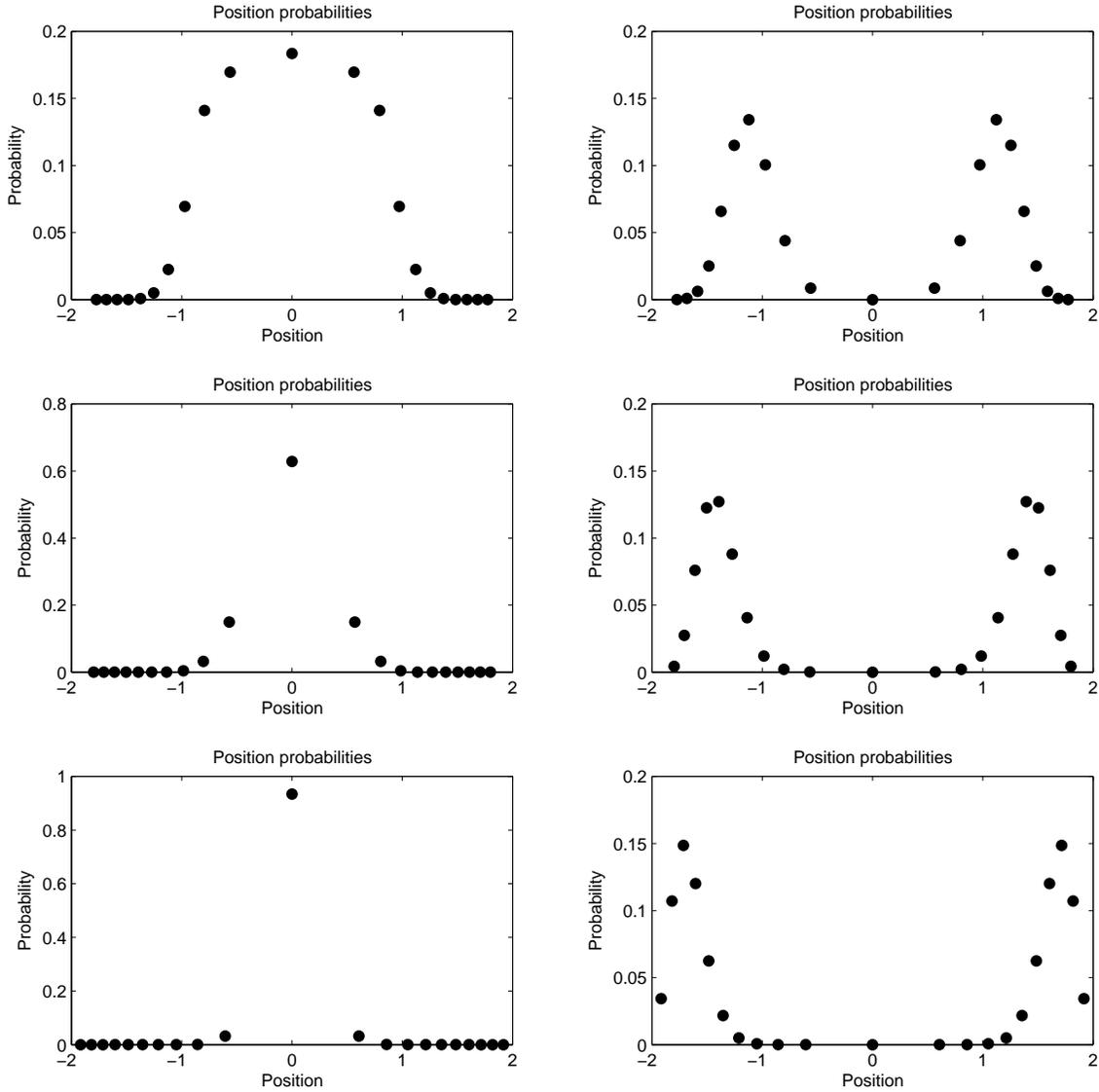}
\caption{Position probability distribution function for the position operator $\hat q_1$,
in the periodic boundary case, when $n=4$ and $p=10$. In the three rows,
$c=0.1$, $c=0.4$, $c=0.8$.
The plotted value is $P(\theta,s,1,\pm x_K)$ for each of the 21 eigenvalues
$\pm x_K$ ($K=0,1,\ldots,10$) of $\hat q_1$. 
This is given for the case when the system is in the stationary state $w(\theta;s)$ corresponding
to the ground state  (minimum energy) in the left column  and in the right column when it is in the 
stationary state $w(\theta;s)$ corresponding to the most excited state (maximum energy).
} \label{PPD-P}
\end{center}
\end{figure}

Let us first consider the case of periodic boundary conditions. 
The position probabilities are independent of $r$, as the system is completely symmetric,
so let us take $r=1$.
We will plot the values 
\[
P(\theta,s,1, \pm x_K), \qquad K=0,1,\ldots,p
\]
for certain values of $\theta$ and $s$. In other words we plot
the position probability distribution function when the system is in a fixed stationary
state $w(\theta;s)$.
Let us consider an explicit example, say $n=4$ (four coupled oscillators) and $p=10$
(so each position operator has 21 distinct eigenvalues). We will plot the position probability distributions
for the ground state (this is the state $w(0;0,p,0,0)=w(0;0,10,0,0)$) and for
the most excited state (this is the state $w(1;0,0,0,p-1)=w(1;0,0,0,9)$).
These distributions are given in Figure~\ref{PPD-P}, for some $c$-values.

Let us make a number of observations on these distributions.
When the system is in the ground state, the probability distribution function of each 
position operator is symmetric around its equilibrium position and unimodal. Of course it
is also discrete (as we are working in finite-dimensional representations). 
As the coupling constant $c$ increases, the ``peak'' around the equilibrium position
is sharper. 
In other words, a)s the coupling constant becomes larger, the oscillators are more likely
to be close to their equilibrium position when the system is in its ground state.

When the system is in its most excited state, the position probabilities are quite different.
The probability of finding the oscillator in its equilibrium position is zero. On the
other hand, there are certain peaks away from the equilibrium position. As $c$ increases,
these peaks are further away from the equilibrium position.
In other words, as the coupling constant becomes larger, the oscillators are more likely
to be further away from their equilibrium position when the system is in its most excited state.

Note that in this figure one also observes the fact that the range of the
spectrum of the position operators is dependent on the coupling constant $c$.
When talking about probabilities of being further or closer to the
equilibrium position we regard this relative to the discrete spectrum 
of $2p+1$ values 
(the middle one being the equilibrium position).

\begin{figure}
\begin{center}
\includegraphics[width=18cm]{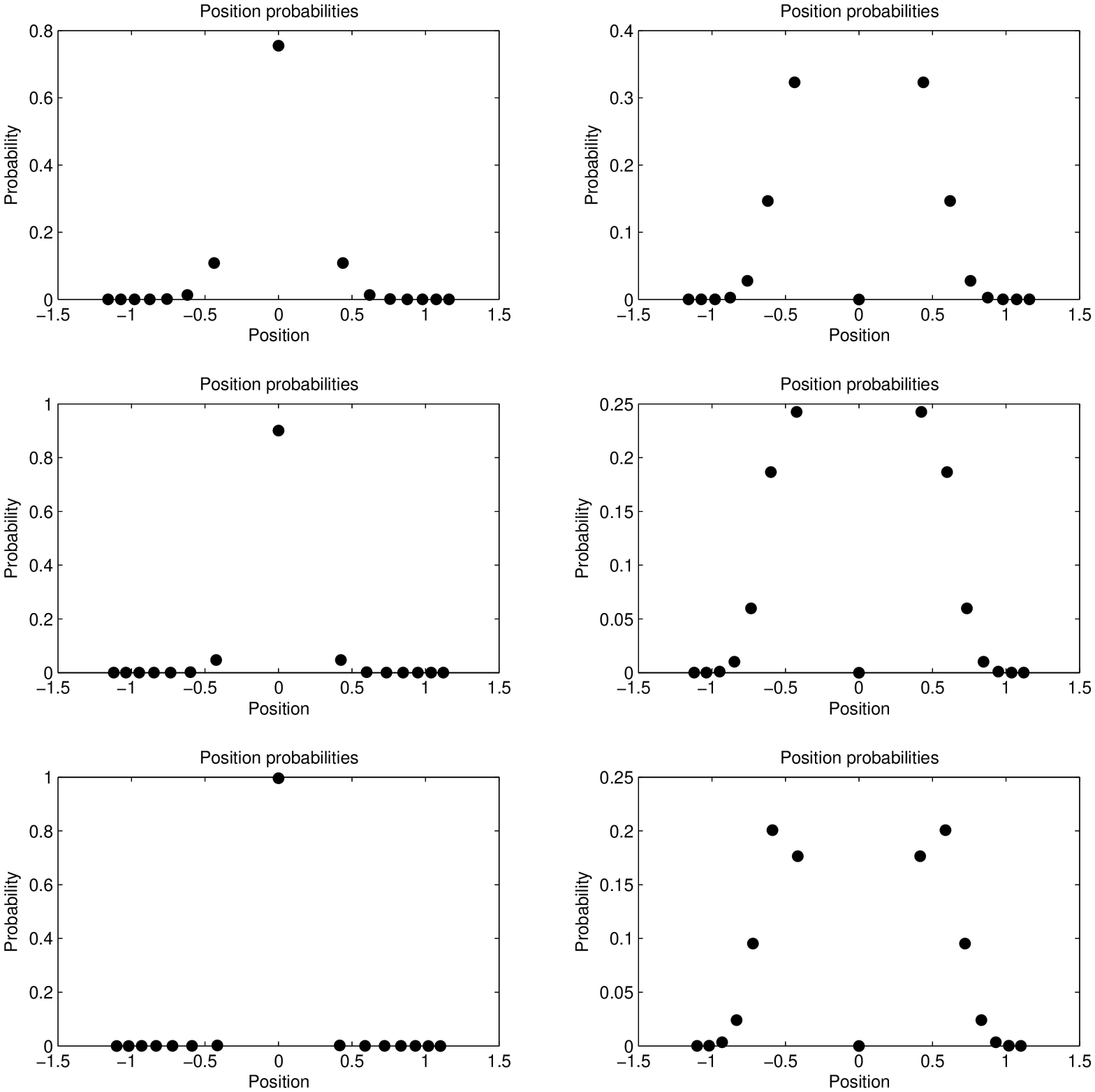}
\caption{Position probability distribution function for the position operator $\hat q_1$,
in the fixed wall boundary case, when $n=6$ and $p=7$. In the three rows,
$c=0.05$, $c=0.2$,  and $c=0.4$.
The plotted value is $P(\theta,s,1,\pm x_K)$ for each of the 15 eigenvalues
$\pm x_K$ ($K=0,1,\ldots,7$) of $\hat q_1$ (next to the wall). 
In the left column, this is given for the case when the system is in the stationary state $w(\theta;s)$ corresponding
to the ground state (minimum energy) and in the right column  for the case when the system is in the 
stationary state $w(\theta;s)$ corresponding to the most excited state (maximum energy).
} \label{PPD-FW1}
\end{center}
\end{figure}

\begin{figure}
\begin{center}
\includegraphics[width=18cm]{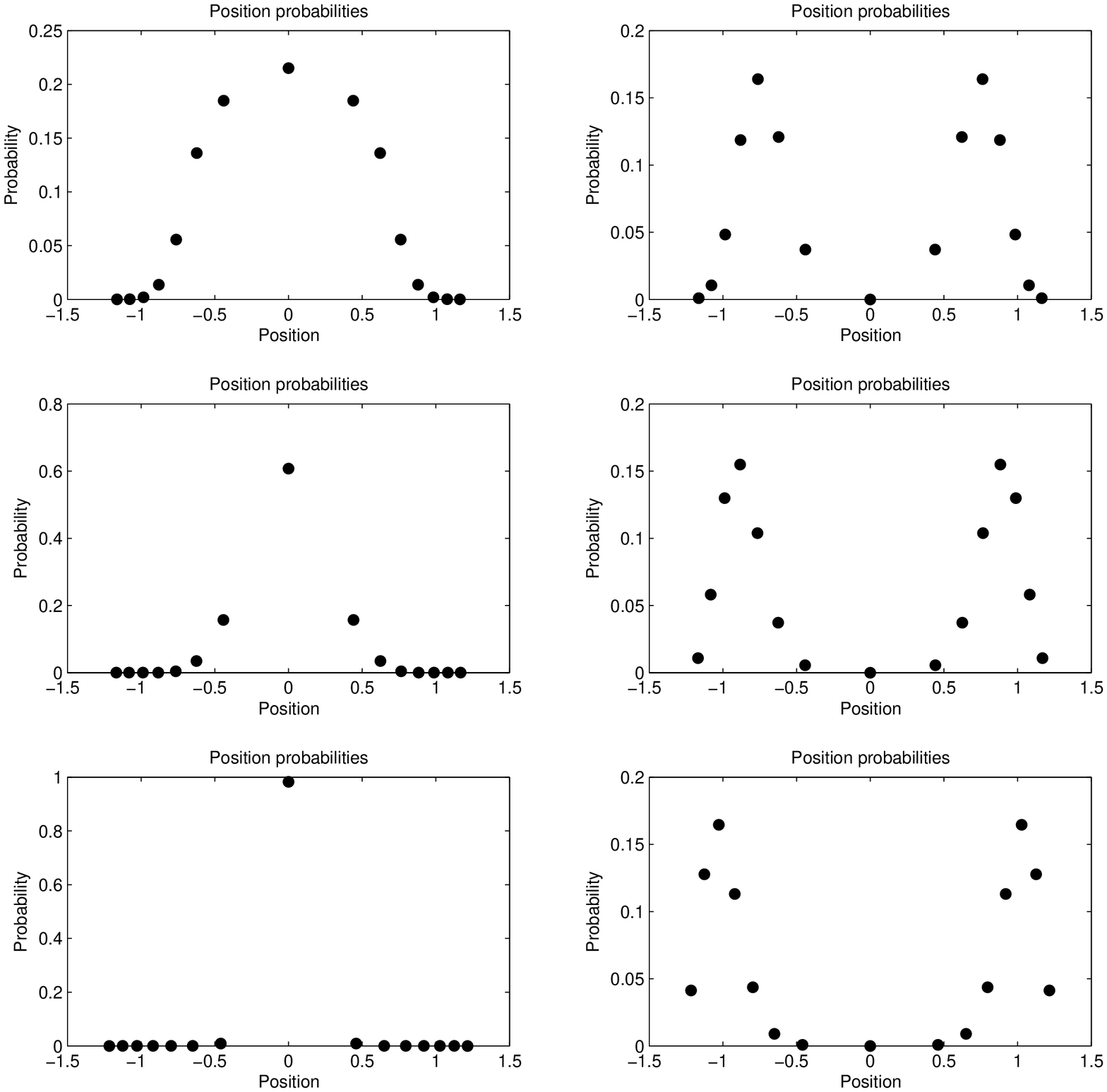}
\caption{Position probability distribution function for the position operator $\hat q_3$,
in the fixed wall boundary case, when $n=6$ and $p=7$. In the three rows,
$c=0.05$, $c=0.2$,  and $c=0.4$.
The plotted value is $P(\theta,s,1,\pm x_K)$ for each of the 15 eigenvalues
$\pm x_K$ ($K=0,1,\ldots,7$) of $\hat q_1$ (next to the wall). 
In the left column, this is given for the case when the system is in the stationary state $w(\theta;s)$ corresponding
to the ground state (minimum energy) and in the right column  for the case when the system is in the 
stationary state $w(\theta;s)$ corresponding to the most excited state (maximum energy).
} \label{PPD-FW3}
\end{center}
\end{figure}

Let us now consider the case of fixed wall boundary conditions. 
The situation is rather different, as the position probabilities 
are depending on $r$. To see the $r$-dependence, we will plot
position probability distribution functions for $r=1$ (the oscillator just
next to the fixed wall) and for $r=3$ (an oscillator away from the wall).
Again, we will plot the values 
\[
P(\theta,s,r, \pm x_K), \qquad K=0,1,\ldots,p
\]
for certain values of $\theta$ and $s$, i.e.\ when the system is in a fixed stationary
state $w(\theta;s)$.
As an explicit example, take $n=6$ (six coupled oscillators) and $p=7$
(so each position operator has 15 distinct eigenvalues). 
We will again plot the position probability distributions
for the ground state (this is the state $w(0;0,0,0,0,0,p)=w(0;0,0,0,0,0,7)$) and for
the most excited state (this is the state $w(1;p-1,0,0,0,0,0)=w(1;6,0,0,0,0,0)$).
These distributions are given in Figure~\ref{PPD-FW1} for $r=1$
and in Figure~\ref{PPD-FW3} for $r=3$, for a number of $c$-values.

Let us again make a number of observations on these distributions.
When the system is in the ground state, the probability distribution function of each 
position operator is symmetric around its equilibrium position and unimodal.
For fixed $c>0$, if the oscillator is closer to the wall ($r=1$) the peak of the
distribution function around the equilibrium position is sharper than for an oscillator
further away from the wall ($r=3$). In other words, the oscillators close to the
wall are closer to their equilibrium position than those further away from the wall.
As the coupling constant $c$ increases, the ``peak'' around the equilibrium position
also becomes sharper, both for $r=1$ and $r=3$. 
In other words, as the coupling constant becomes larger, the oscillators are more likely
to be close to their equilibrium position.

When the system is in its most excited state, the position probabilities are rather different.
The probability of finding the oscillator in its equilibrium position is zero. 
For $c>0$, there are certain peaks away from the equilibrium position, both for $r=1$ and for $r=3$. 
Close to the wall ($r=1$), these peaks are closer to zero than away from the wall ($r=3$).
So also in this most excited state, the oscillators close to the wall ``oscillate less heavily''
than those away from the wall. 
As $c$ increases, these peaks are further away from the equilibrium position, both
for $r=1$ and for $r=3$.
So the oscillators are more likely
to be further away from their equilibrium position when the coupling constant increases.

\setcounter{equation}{0}
\section{Conclusions}

We have examined properties of noncanonical solutions of two quantum systems:
chains of coupled harmonic oscillators with periodic boundary conditions or with fixed wall
boundary conditions.
These new solutions arise from an approach as a WQS, allowing more classes of solutions
than just the canonical one.

For the solutions examined here, the position and momentum operators are (odd) elements of the Lie
superalgebra $\gl(1|n)$. The physical properties of the system then follow from the 
representations of $\gl(1|n)$ considered. Here, we have introduced the rather simple ladder 
representations $V(p)$, a class of unitary irreducible representations of $\gl(1|n)$ (or
rather, of its compact form $\u(1|n)$). 
For these representations, we have determined the energy spectrum, which was rather easy
due to the simple action of the Hamiltonian operator in the standard basis of $V(p)$.
We have also determined the spectrum of the position operators; this task was more difficult
because of the more complicated action of these operators in the basis of $V(p)$.
The techniques developed in~\cite{LSV07} allow to construct explicitly the eigenvectors
of the position operators, for both systems under consideration.

The analysis of the spectrum of the position operators, and their probabilities when the
system is in a certain stationary state, lead to interesting properties.
The spectrum is discrete, centered around the equilibrium position; the number of
possible position values depends on $p$ (it is $2p+1$). The width of this discrete
support depends on the coupling constant $c$, and -- only in the case of fixed wall
boundary conditions -- on the order $r$ of the oscillator in the chain of $n$ oscillators.
The position probability distributions, discussed in detail in the previous section,
have properties similar to those of a classical system of coupled oscillators.

The current paper was still dealing with solutions arising from the Lie superalgebra
$\gl(1|n)$, although a different class of representations was considered than in~\cite{LSV06}. 
As indicated in a previous paper~\cite{LSV06}, also other types of solutions for the compatibility
conditions exist, for example in terms of the Lie superalgebra $\osp(1|2n)$.
It would be interesting to investigate other solutions in terms of this
orthosymplectic Lie superalgebra, even though the analysis is expected to be
rather difficult.

\section*{Acknowledgments}
N.I.\ Stoilova was supported by a project from the Fund for Scientific Research -- Flanders (Belgium)
and by project P6/02 of the Interuniversity Attraction Poles Programme (Belgian State -- 
Belgian Science Policy).
S.\ Lievens was also supported by project P6/02.

\end{document}